\RequirePackage{lineno}
\documentclass[twocolumn,letterpaper,aps,prl,longbibliography,superscriptaddress,showpacs,floatfix]{revtex4-1}

\usepackage{graphicx}	

\usepackage{xspace}	

\usepackage{hyperref}

\begin{document}

\title{Global analysis of color fluctuation effects in proton-- and
  deuteron--nucleus collisions at RHIC and the LHC}

\author{M. Alvioli}
\affiliation{Consiglio Nazionale delle Ricerche, Istituto di Ricerca per la Protezione Idrogeologica,
  via Madonna Alta 126, I-06128 Perugia, Italy}
\author{L. Frankfurt }
\affiliation{Tel Aviv University, Tel Aviv,  Israel}
\affiliation{104 Davey Lab, The Pennsylvania State University,
  University Park, PA 16803, USA}
\author{D.V.~Perepelitsa}
\affiliation {University of Colorado, Boulder, CO 80309
 USA}
\author{M. Strikman}
\affiliation{104 Davey Lab, The Pennsylvania State University,
  University Park, PA 16803, USA}

\date{\today}

\begin{abstract}
We test the hypothesis that configurations of a proton with a
large-$x$ parton, $x_p \gtrsim 0.1$, have a smaller than average
transverse
size. The 
application of the 
QCD $Q^2$ evolution equations 
shows
that these small configurations also have a significantly smaller
interaction strength, which has observable consequences in proton --
nucleus collisions.  We perform a global analysis of jet production
data in proton-- and deuteron--nucleus collisions at RHIC and the
LHC. Using a model which takes a distribution of interaction strengths
into account, we quantitatively extract the $x_p$-dependence of the
average interaction strength, $\sigma(x_p)$, over a wide kinematic
range. By comparing the RHIC and LHC results, our analysis finds that
the interaction strength for small configurations, while suppressed,
grows faster with collision energy than does that for average
configurations. We check that this energy dependence is consistent
with the results of a method which, given $\sigma(x_p)$ at one energy,
can be used to quantitatively predict that at another. This finding
further suggests that at even lower energies, nucleons with a
large-$x_p$ parton should interact much more weakly than those in an
average configuration, a phenomenon in line with explanations of the
EMC effect for large-$x_p$ quarks in nuclei based on color screening.
\end{abstract}


\maketitle


Hadrons are composite, quantum mechanical systems with a varying
spatial and momentum configuration of their internal quark and gluon
constituents. In sufficiently high energy processes, these
configurations remain approximately fixed during the time of the
collision.  Thus certain physical properties of the parton system of a
rapidly moving hadron, such as the total transverse area occupied by
the color fields, may change collision by collision, a phenomenon we
refer to as \textit{color
  fluctuations}~\cite{Alvioli:2013vk,Alvioli:2014eda}. These
variations in the internal structure of hadrons have a wide range of
observable consequences, such as inelastic
diffraction~\cite{Pomeranchuk1953, Feinberg1956,
  Frankfurt:2013ria}. In quantum chromodynamics (QCD), the
configurations in which a large ($>10$\%) fraction of the hadron's
momentum is carried by a single parton are spatially compact. For
these cases, in the wide range of energies where non-linear
(saturation) effects are expected to be
small~\cite{Mclerran:2013mza,Alvioli:2012ba}, the interaction strength
of the entire configuration decreases along with the overall area
occupied by color (for a review and references see
Ref.~\cite{Frankfurt:1994hf}).  Furthermore, while the interaction
strength for such small configurations is reduced overall, it rises
rapidly with collision energy due to a fast increase of the gluon
density at small $x$.  In this paper, we quantitatively investigate
these properties of QCD systems in proton-- and deuteron--nucleus
($p/d$+A) collision data at the Large Hadron Collider (LHC) and the
Relativistic Heavy Ion Collider (RHIC), respectively.

\begin{figure}[!t]
\includegraphics[width=1.0\linewidth]{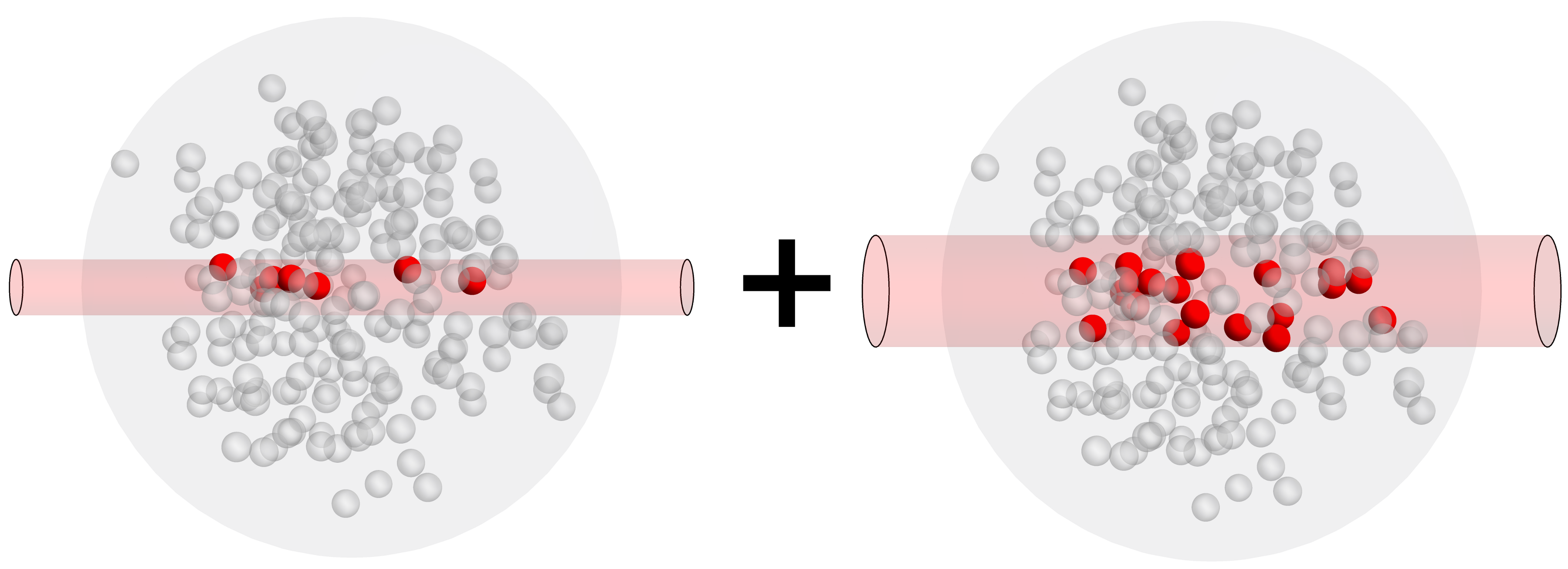}
\caption{\label{fig:fig0} Schematic representation of a
  proton--nucleus collision with a fixed geometry of the target nucleus,
  with a more weakly (more strongly) interacting projectile proton on
  the left (right). The red tube shows the projection of the
  projectile proton's transverse size through the nucleus, with
  impacted nucleons in red. Typical observables have contributions
  from both types of events, while large-$x_p$ configurations may
  preferentially select weakly interacting cases (left).}
\end{figure}

Fig.~\ref{fig:fig0} symbolically illustrates how proton configurations
of two different sizes contribute to $p$+A interactions. For many
processes, a large number of projectile configurations contribute to a
studied observable, resulting in a lack of sensitivity to color
fluctuation effects. However, in processes to which only a restricted
subset of projectile configurations contribute, these effects are
important for understanding the experimental data. Historically, they
have played a role in interpreting multiplicity distributions in
nuclear collisions~\cite{Heiselberg:1991is} and in describing the
coherent diffractive production of
dijets~\cite{Frankfurt:1993it,Frankfurt:2000jm,Aitala:2000hc}\footnote{The
  first theoretical study of the coherent dissociation of pions
  scattering off nuclei into two jets was performed in
  Ref.~\cite{Bertsch:1981py}. The obtained result contradicts the QCD
  factorization theorem~\cite{Frankfurt:1993it,Frankfurt:2000jm} and
  A-dependence and $p_t$ dependence of the process measured in
  data~\cite{Aitala:2000hc}.}.

Experimentally, $p$+A collisions with a restricted subset of
projectile configurations may be selected with a special trigger such
as a hard QCD or electroweak process involving a large-$x_p$ ($\gtrsim
0.1$) parton in the proton~\cite{Frankfurt:1985cv}. In these
large-$x_p$ configurations, color charge screening within the dominant
Feynman diagrams suppresses the gluon field and density of $q\bar{q}$
pairs, leading to an interaction cross-section which is smaller but
grows rapidly with energy (for a review of this phenomenon in HERA
data, see Ref.~\cite{Abramowicz:1998ii}). 

The success of the quark counting
rules~\cite{Brodsky:1973kr,Brodsky:1974vy} indicates what chain
diagrams dominate at large $x$. Analysis of these Feynman diagrams
\cite{Frankfurt:1981mk} indicates that quark transverse momenta should
be rather large and hence the 3q configurations should have size much
smaller than average.  ~\footnote{However, an opposite trend is
  present for the relativistic Gaussian wave function considered
  e.g. in \cite{Brodsky:2018vyy}.
   For example, consider a meson
  with wave function $\psi^2(x,k_t) \propto \exp(-c \cdot
  k_t^2/(x(1-x))$. The transverse area occupied by a large $x$
  configuration grows as $\propto 1/[x(1-x)]$. At the same time, in
  the quark models which feature a singular short distance potential,
  such large-$x$ configurations do shrink.}.

In $p$+A collisions, the shrinking of the proton configuration in
large-$x_p$ scattering events should lead to a decrease in the average
number of nucleon--nucleon ($NN$) interactions between the projectile
and target nucleus, $\nu$, relative to that for collisions with a more
typical proton configuration. In the $p$+A case, $\nu$ also coincides
with the number of wounded nucleons in the target nucleus. This
feature should also be present in $d$+A collisions, although the
magnitude of the effect is diminished due to the unaffected nucleon in the 
deuteron contributing with an average over its configurations. $\nu$ is indirectly 
measured in experiments via the soft particle
multiplicity~\cite{Adare:2013nff,Aad:2015zza,Adam:2014qja}.

Measurements which can test these properties of QCD were recently
performed in proton--lead ($p$+Pb) collisions at the LHC
\cite{ATLAS:2014cpa,Chatrchyan:2014hqa} and deuteron--gold ($d$+Au)
collisions at RHIC \cite{Adare:2015gla} at center of mass energies of
$5.02$~TeV and $200$~GeV, respectively. In these data, the production
of large transverse momentum ($p_t$) jets was studied in the
large-$x_p$ kinematic region as a function of hadronic activity in the
downstream nucleus-going direction ($\eta < -3$). Hadron production
rates in this rapidity range are correlated with $\nu$, and have been
experimentally shown to be insensitive to energy-momentum conservation
effects related to jet production at mid- and forward (proton-going)
rapidities~\cite{Aad:2015ziq} (though such correlations were expected
in some models of the process under
consideration~\cite{Armesto:2015kwa}). Each experiment observed a
qualitatively consistent picture: for events with jets originating
from a large-$x_p$ scattering, the geometric (eikonal) model strongly
underestimates the number of events with low hadronic activity
(geometrically ``peripheral'' events in the classical picture) and
overestimates those with a large hadronic activity (``central''
events). However, inclusive jet production rates were unmodified,
$\sigma^{p+A} = A \sigma^{p+p}$, as expected from QCD factorization
and the small modification of the nuclear parton densities in this
region~\cite{Armesto:2015lrg}.
  
In our previous analysis~\cite{Alvioli:2014eda} we demonstrated that
color fluctuation effects which led to a more weakly interacting
large-$x_p$ configuration could quantitatively describe the ATLAS data
for jet production at $x_p \approx 0.6$. In this paper, we present a
unified analysis of ATLAS~\cite{ATLAS:2014cpa} and
PHENIX~\cite{Adare:2015gla} data to study the collision energy and
$x_p$-dependence of this effect in detail. CMS has observed a
qualitatively similar effect in the centrality dependence of di-jet
production~\cite{Chatrchyan:2014hqa}. However, those data are
presented with an open $p_t$ selection and as a function of the system
pseudorapidity $(\eta_1 +\eta_2)/2$, and thus integrate over a rather
wide distribution of $x_p$ values. Thus we do not include it in the
present extraction, which relies on isolating narrow ranges of $x_p$
values.
 
To model the effects of color fluctuations in $p$+A collisions, we use
the Monte Carlo algorithm developed in
Refs.~\cite{Alvioli:2013vk,Alvioli:2014sba}, of which we summarize the
main features here. In our procedure, the probability distribution,
$P_N(\sigma)$, for a projectile nucleon configuration to have a total
cross-section for an interaction with another nucleon in the target,
$\sigma$, is given by

\begin{equation}
    P_N(\sigma)= C \frac{\sigma}{\sigma\,+\,\sigma_0}\,
    \mathrm{exp} \left\{-\frac{(\sigma/\sigma_0\,-\,1)^2}
           {\Omega^2}\right\}.
    \label{psigma}
\end{equation}  
 Note here that configurations with small $\sigma$ correspond to the color transparency 
 regime which contributes very little to the phenomena we consider here.

The parameters of $P_N$ are determined from analyses of data on
diffractive processes in hadronic collisions, which are sensitive to
the size of the fluctuations, as discussed in
Ref.~\cite{Blaettel:1993rd}. This form consistently accounts for
several expected properties of the projectile hadron wave functions:
(1) it follows from a generalization of the quark counting rules, $P_N
\to 0$ as $\sigma\to 0$; (2) $P_N$ is approximately Gaussian for
$\sigma \sim \sigma_0$; (3) the first two moments of the distribution
give the conservation of probability ($\int P_N d\sigma = 1$) and
define the average total cross-section ($\int P_N \sigma d\sigma =
\sigma_{tot}$); (4) it smoothly interpolates between the expected
behavior at small and large values of $\sigma$ (the former of which is
in the color transparency regime). A different parameterization of
$P_N$ at RHIC energies may be found in
Ref.~\cite{Coleman-Smith:2013rla}, and other approaches based on
fluctuations in the positions of proton constituents are also
discussed in the literature, see e.g.~\cite{Albacete:2016gxu}.

We determine the distribution of $\nu$ values in $p$+A collisions by
extending standard simulation procedures~\cite{Loizides:2014vua} to
include fluctations in the proton interaction strength and other
effects. The spatial configuration of nucleons in the nucleus are
generated according to a Woods-Saxon distribution but taking into
account short-range $NN$ spatial correlations which affect the nuclear
two-body density~\cite{Alvioli:2009ab}. The probability that the
projectile nucleon interacts with a target nucleon varies with their
transverse displacement according to the profile function of the
interaction. In addition, the probability of a hard interaction was
determined through the convolution of generalized parton distributions
(which describe the longitudinal and transverse distributions of
partons) in the projectile proton and target nucleons, as discussed in
Ref.~\cite{Alvioli:2013vk}. Thus, the model takes into account the
spatial localization of hard partons close to the center of the
nucleon~\cite{Frankfurt:2010ea}.

One of the struck nucleons in the target is randomly chosen to contain
the hard scattering, while the remaining nucleons undergo soft
interactions with the inelastic fraction of the fluctuating
cross-section ($\approx 0.75 \sigma_{tot}$). For $d$+A collisions, the
configuration of the deuteron is sampled according to the projection
of its wavefunction into the transverse plane. In this way, the model
provides the distribution over the number of $NN$ interactions, $\nu$,
for $p/d$+A collisions.

\begin{figure*}[!t]
\includegraphics[width=0.24\linewidth]{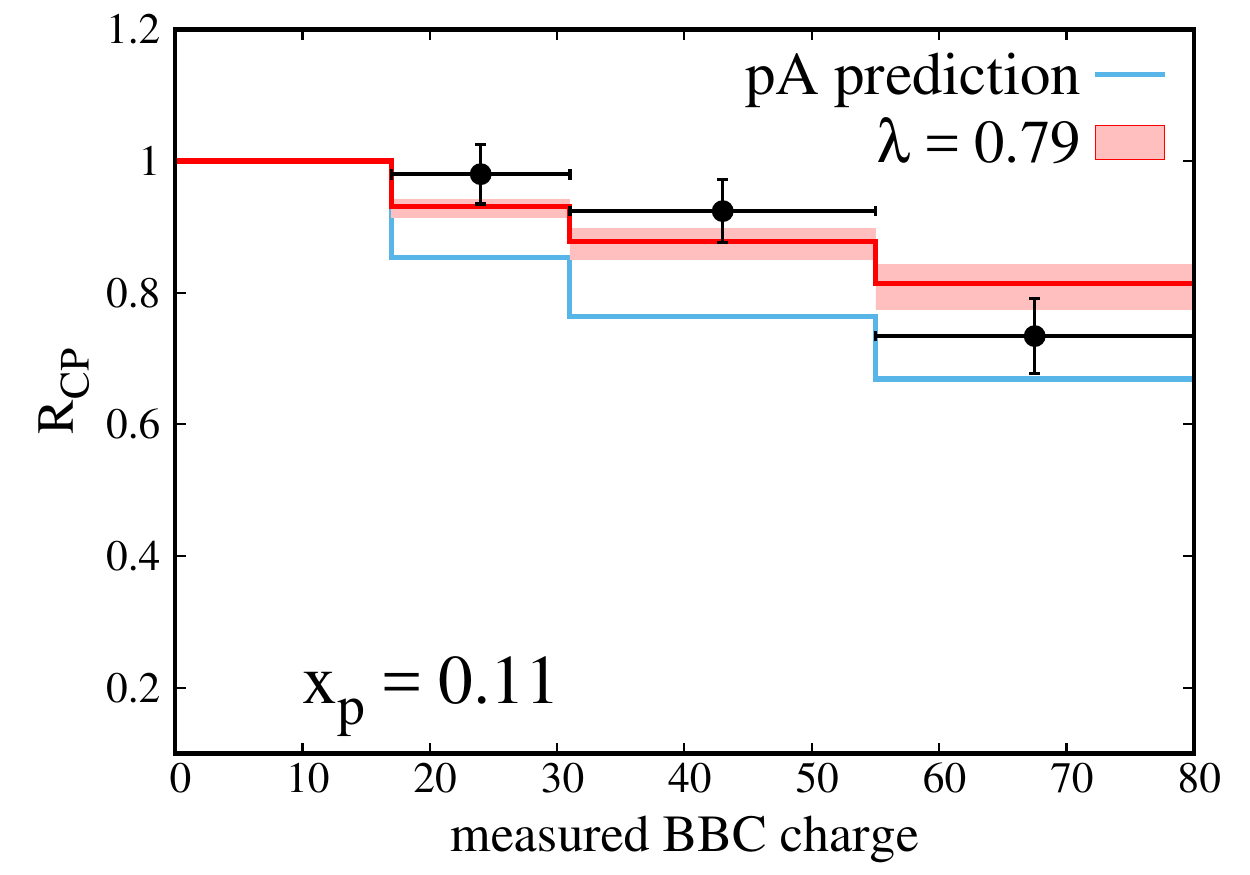}
\includegraphics[width=0.24\linewidth]{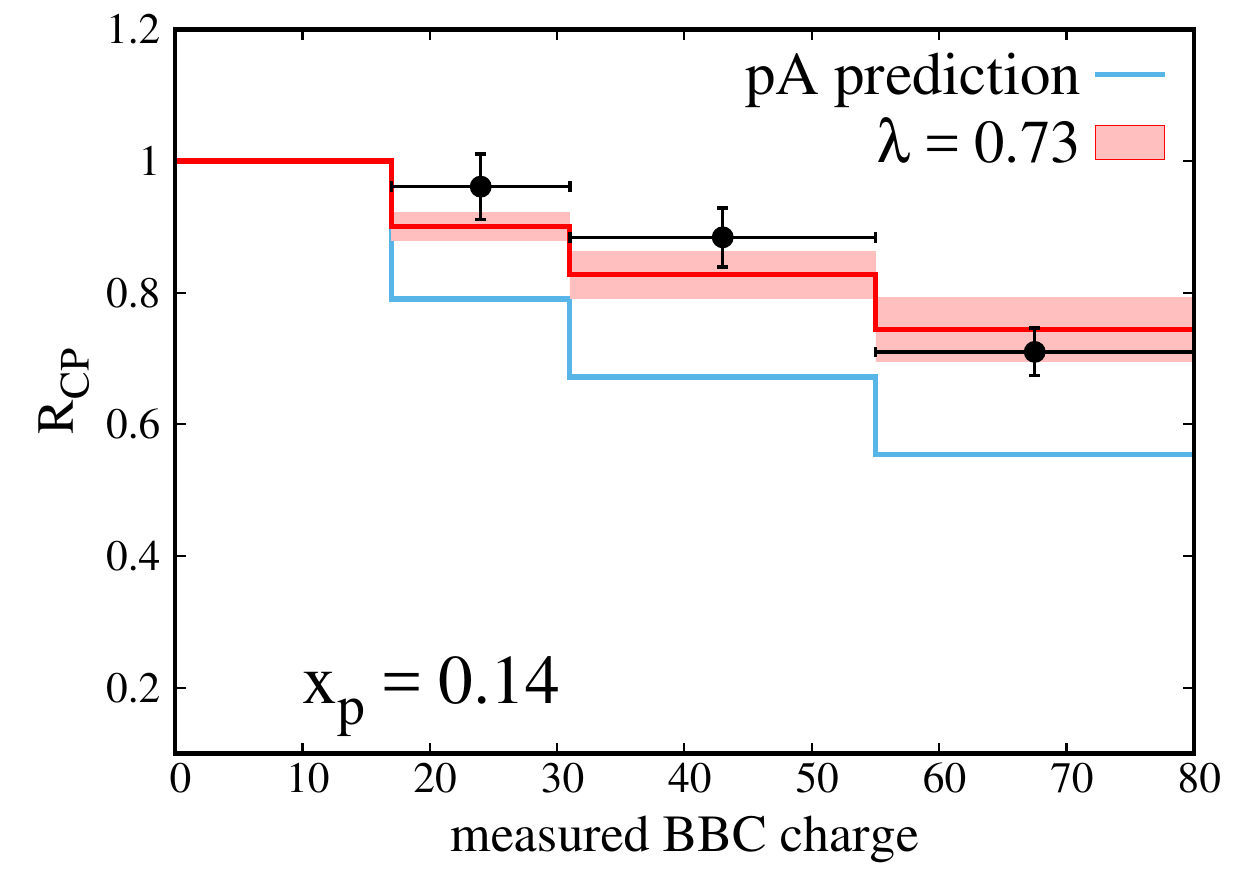}
\includegraphics[width=0.24\linewidth]{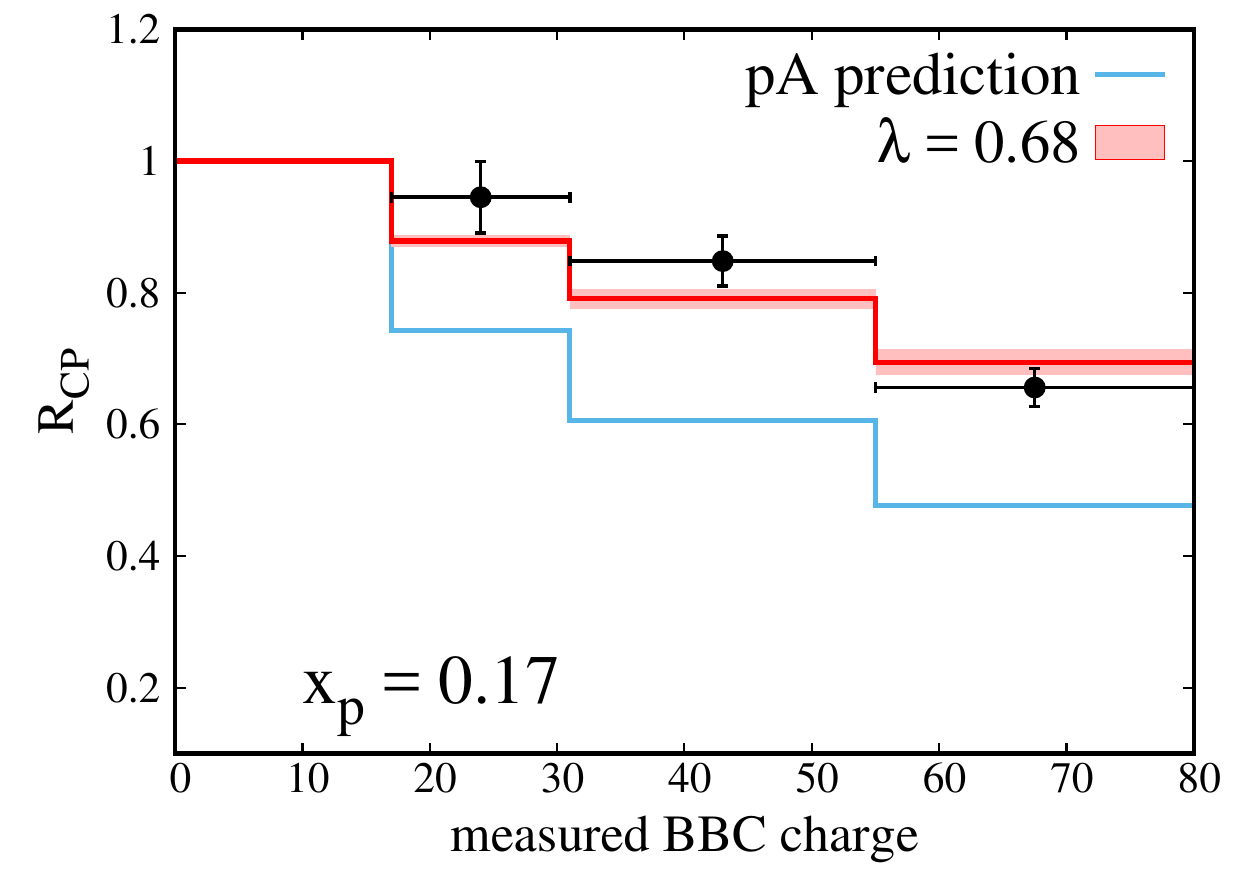}
\includegraphics[width=0.24\linewidth]{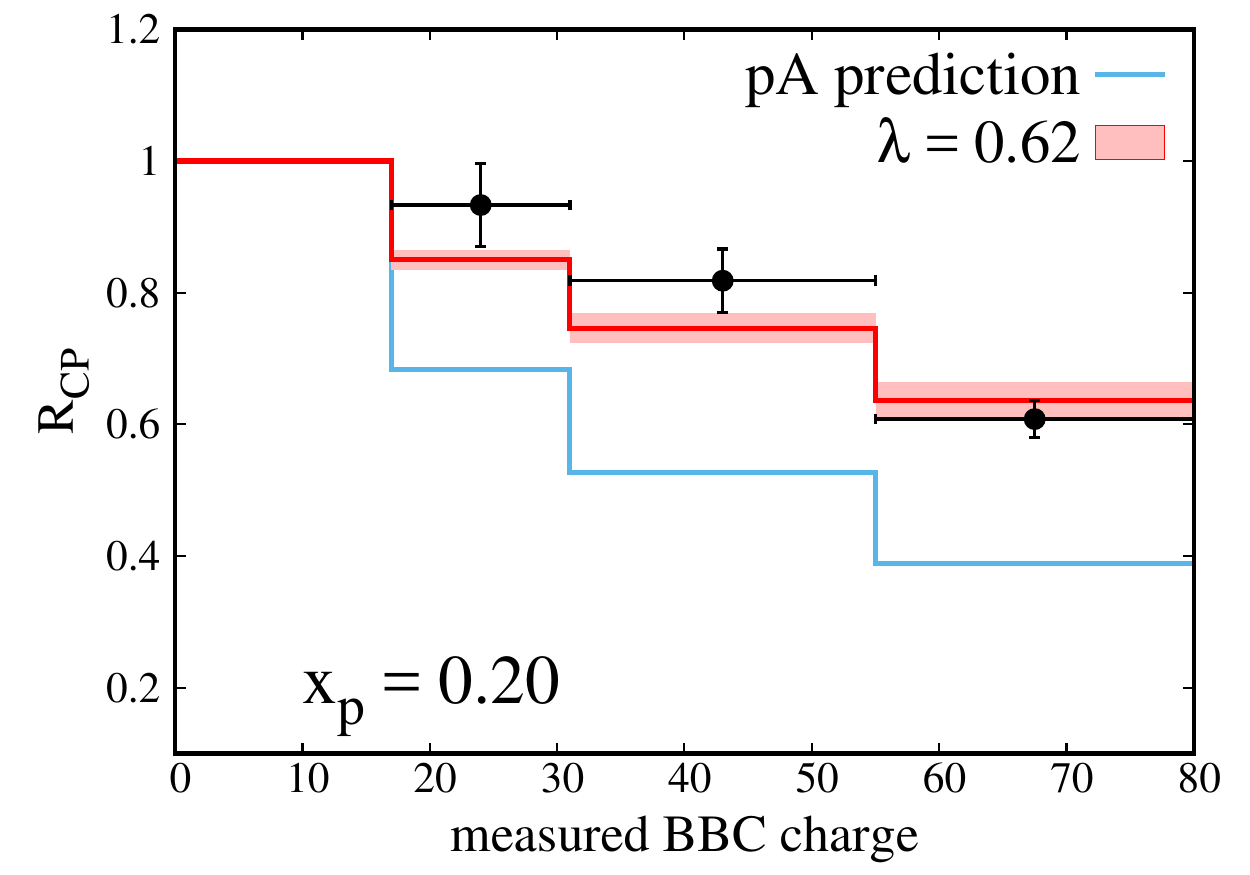}
\includegraphics[width=0.24\linewidth]{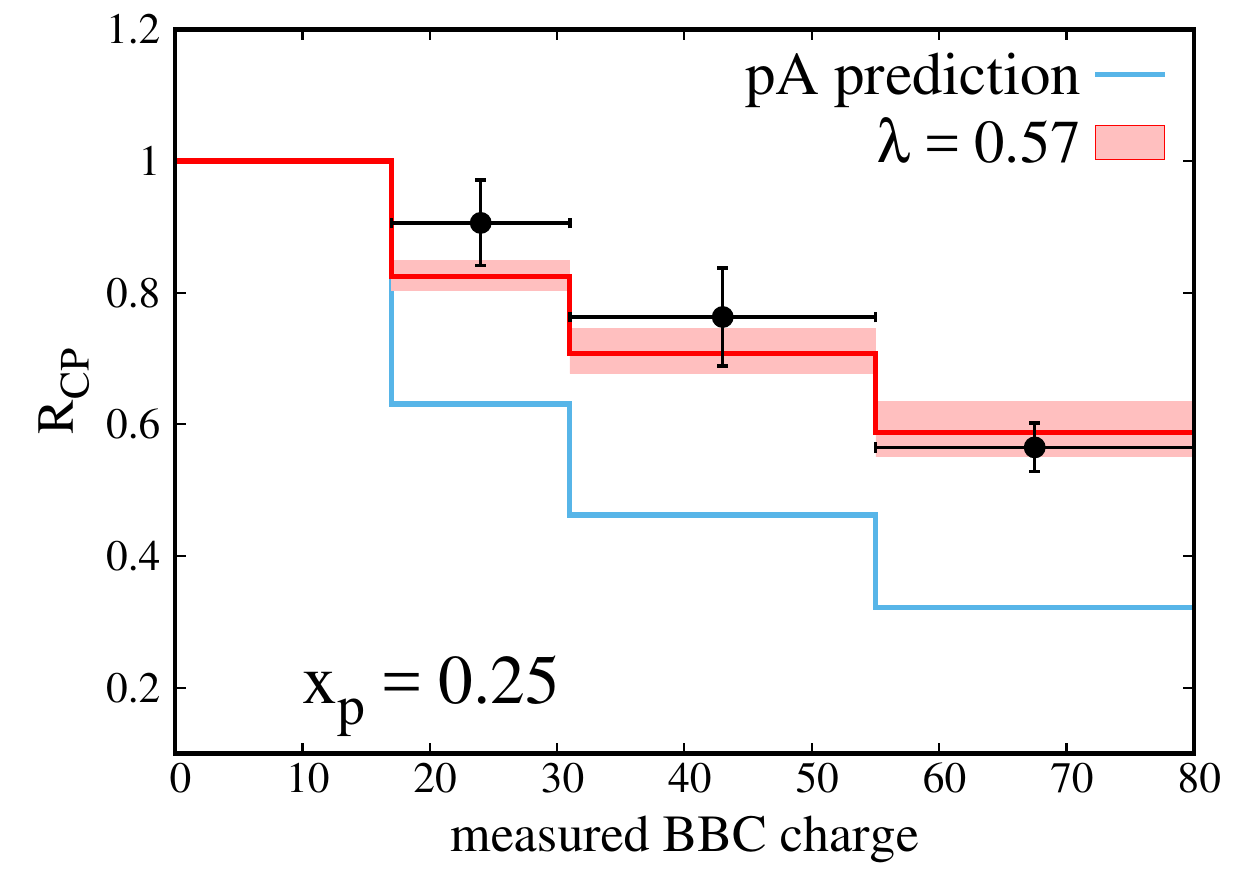}
\includegraphics[width=0.24\linewidth]{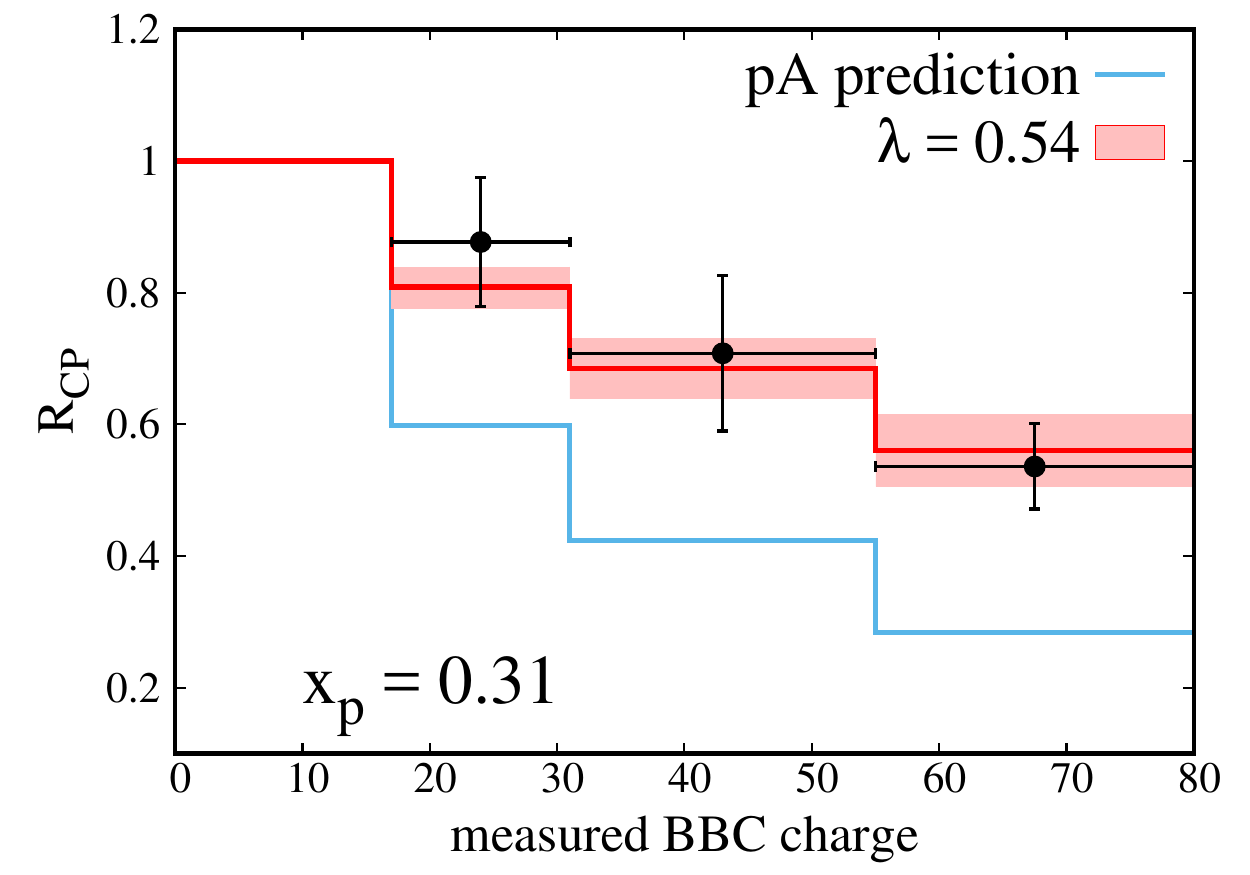}
\includegraphics[width=0.24\linewidth]{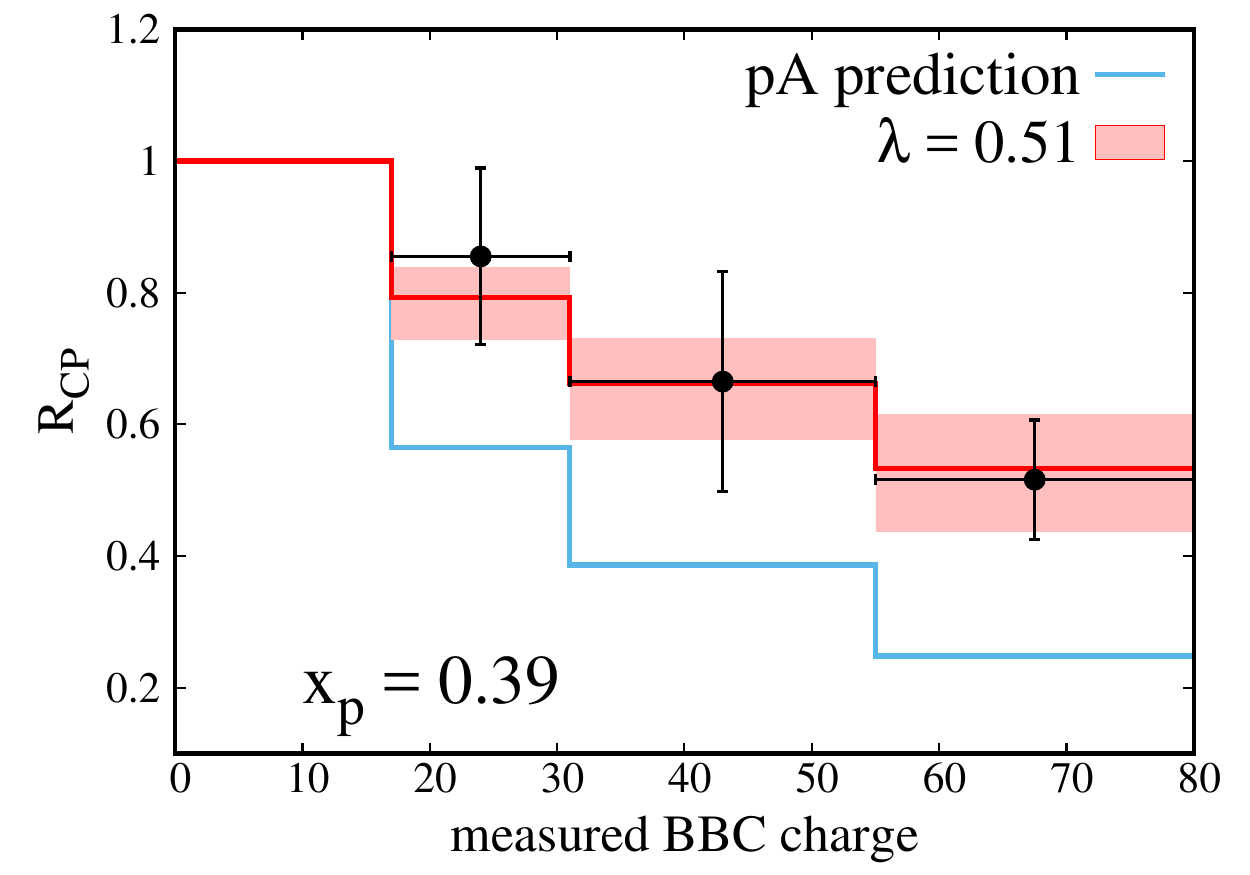}
\includegraphics[width=0.24\linewidth]{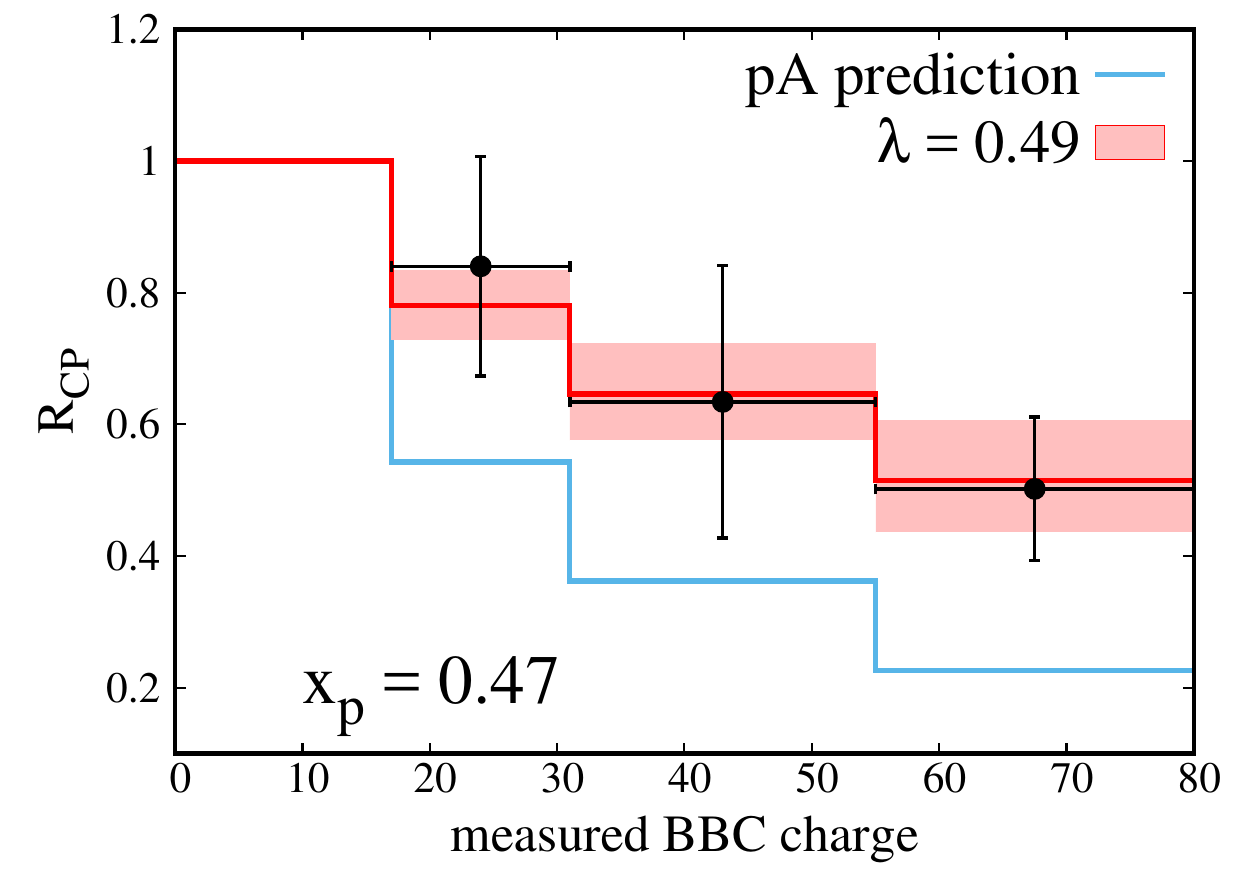}
\caption{\label{fig:fig2} Comparison of the RHIC deuteron--gold
  nuclear modification factor data (black points) in different
  hadronic activity bins, to those in our model (shaded band), and to
  predictions for proton--gold data at RHIC (blue line). Each panel
  shows a different $x_p$ range.}
\end{figure*}

\begin{figure*}[!t]
\includegraphics[width=0.24\linewidth]{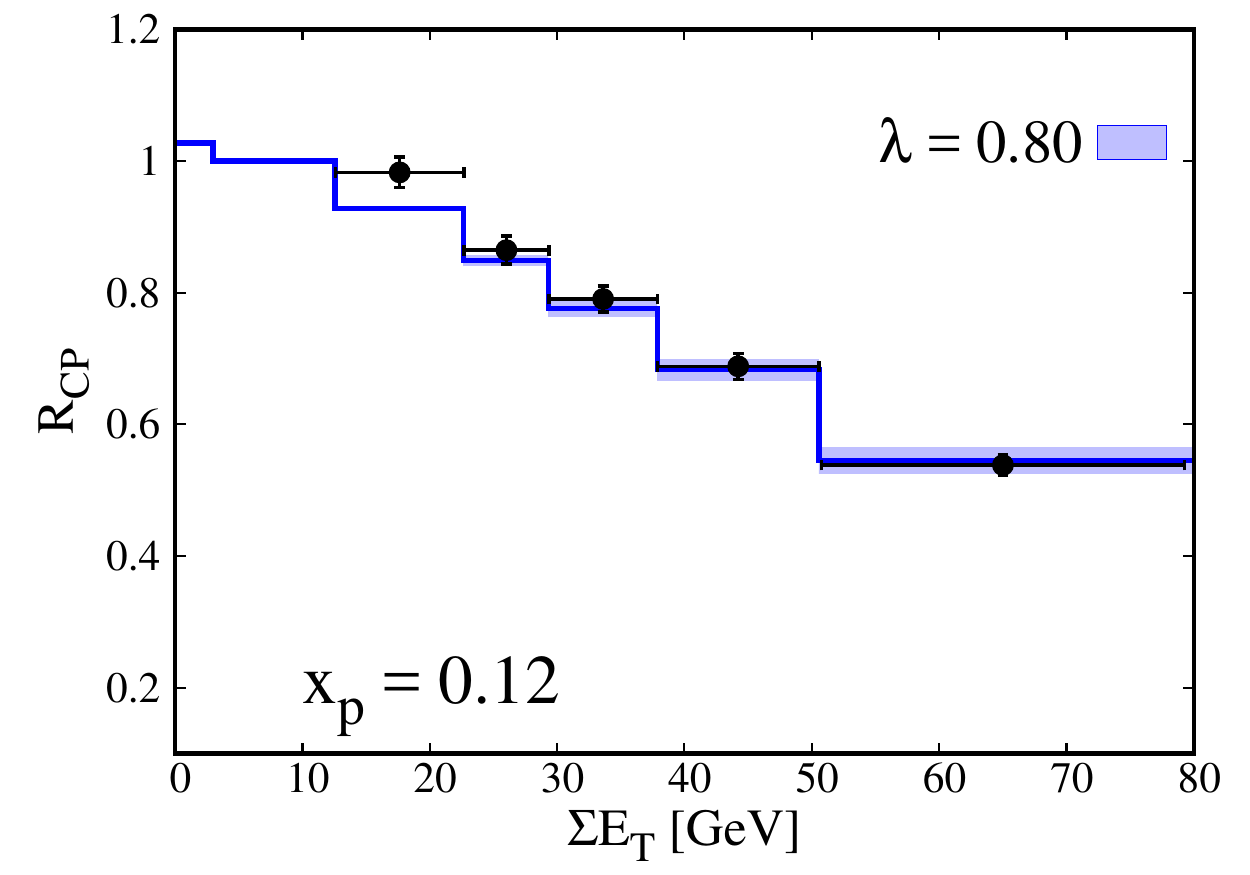}
\includegraphics[width=0.24\linewidth]{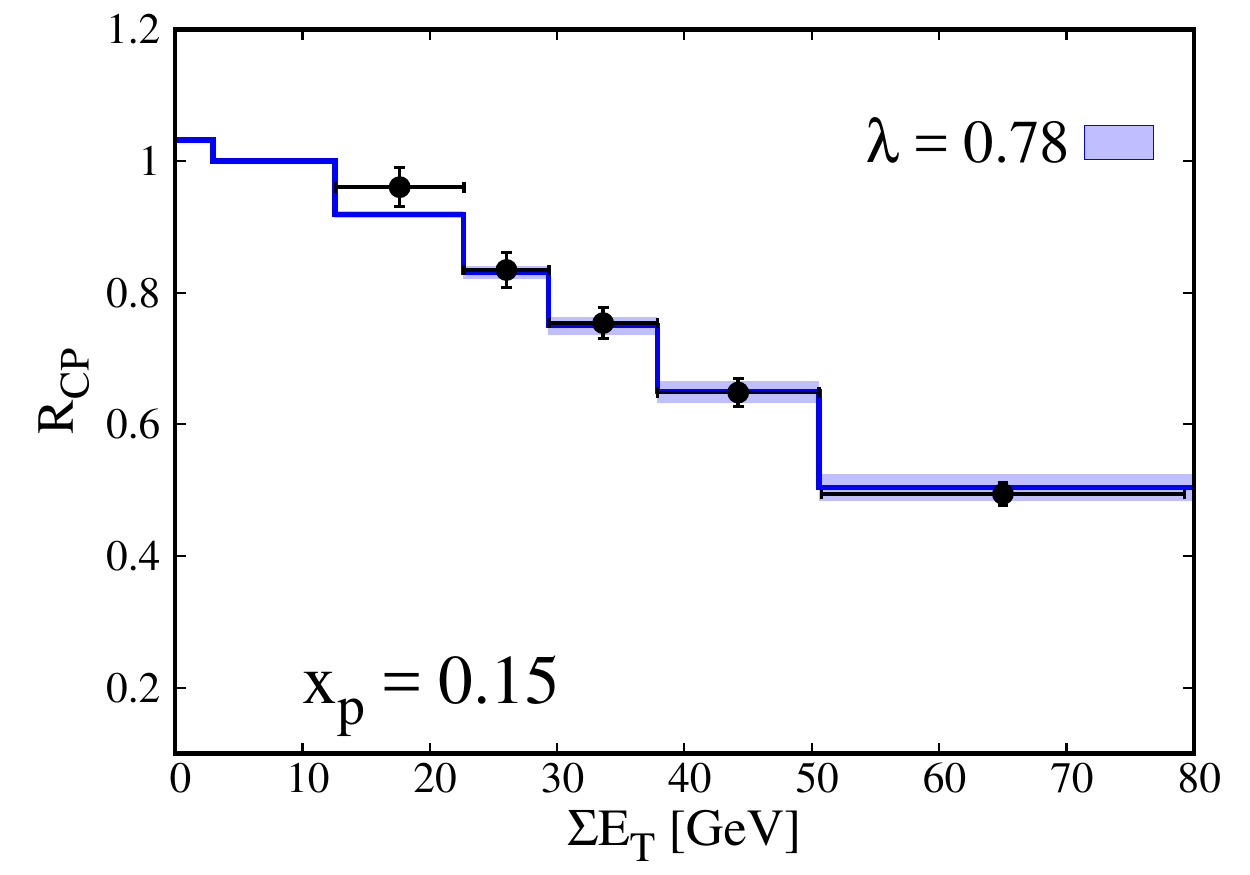}
\includegraphics[width=0.24\linewidth]{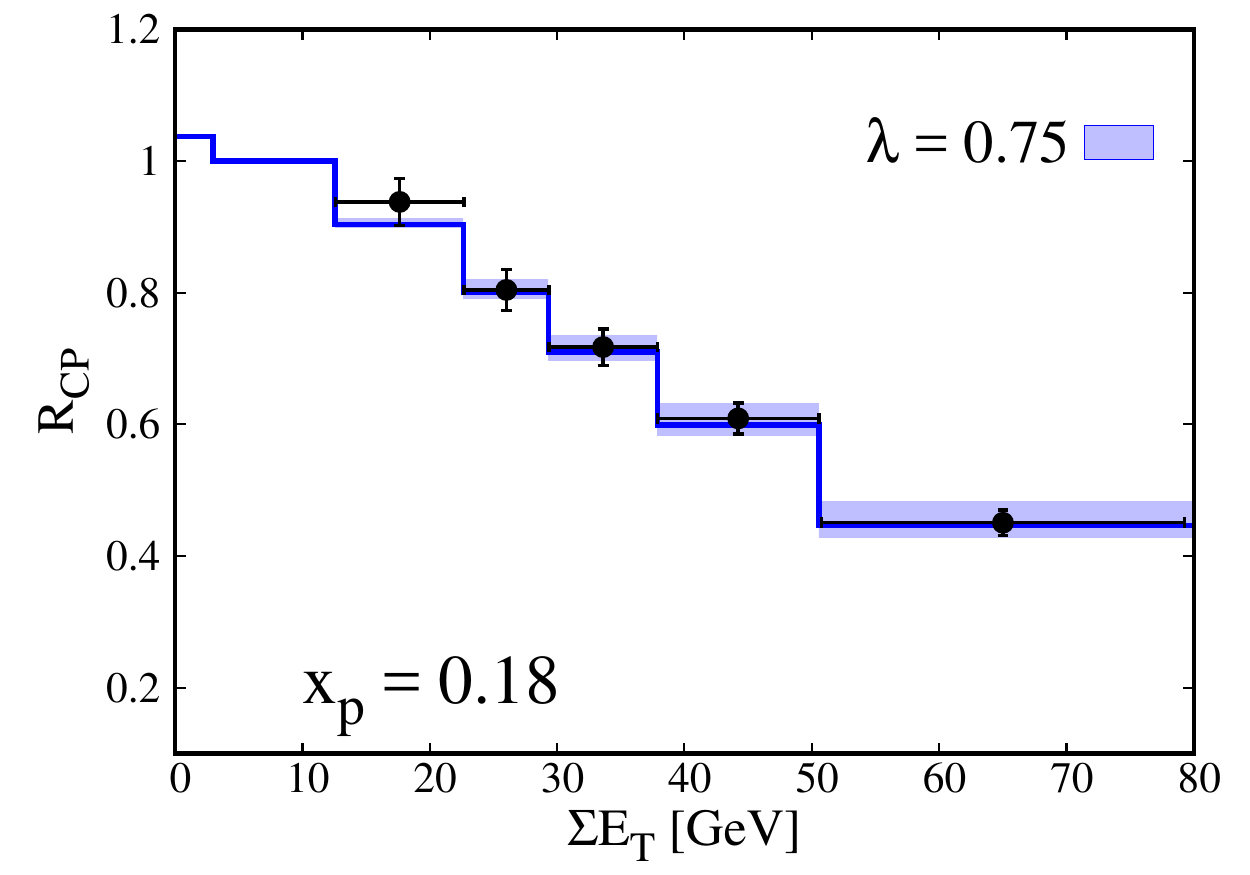}
\includegraphics[width=0.24\linewidth]{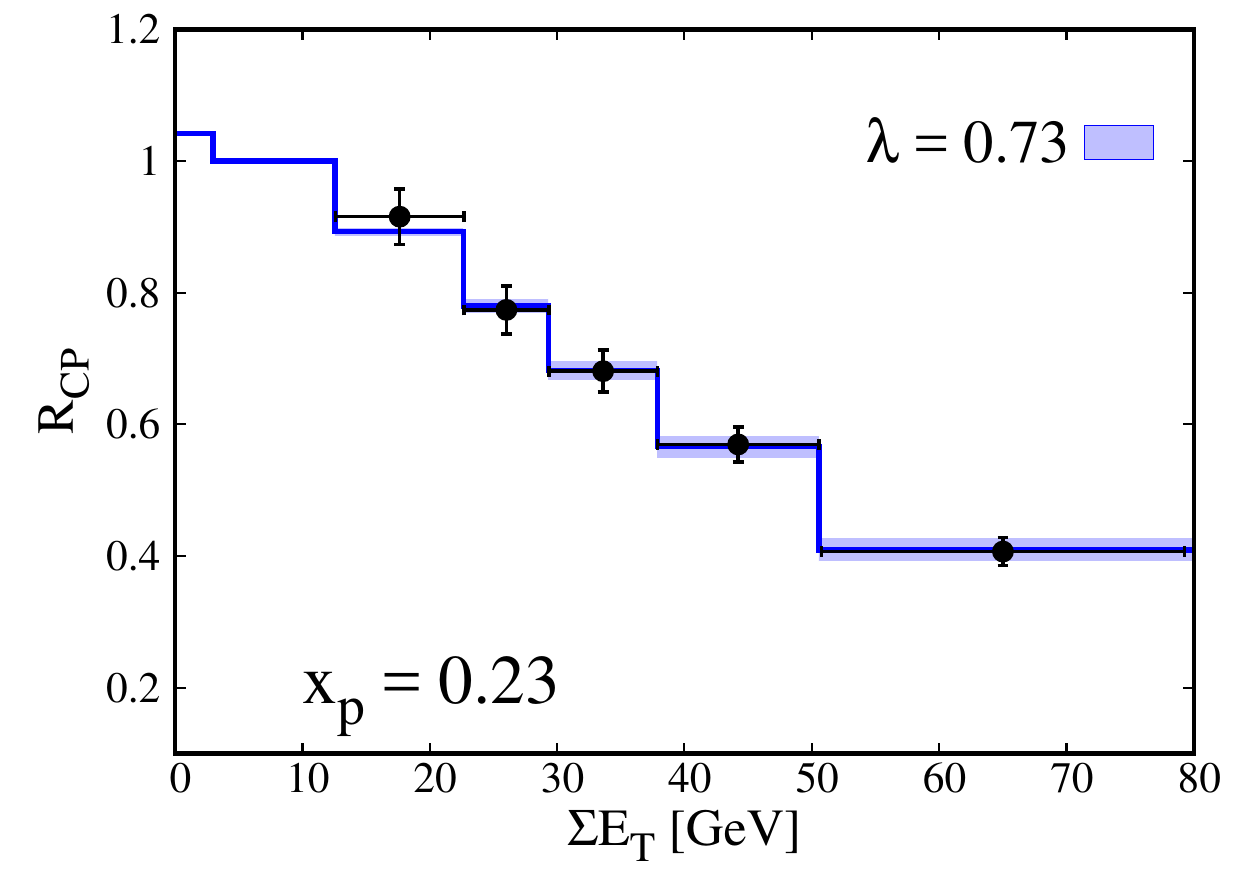}
\includegraphics[width=0.24\linewidth]{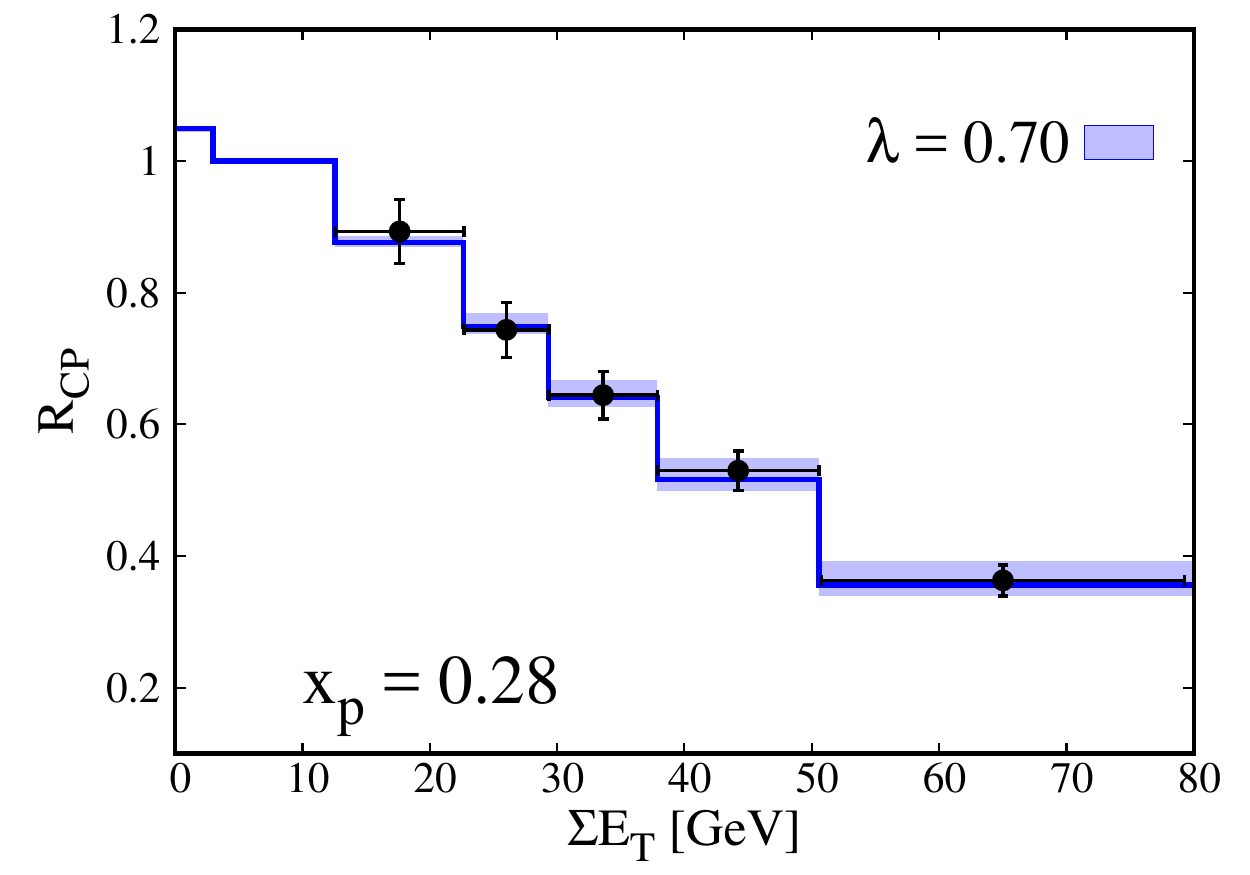}
\includegraphics[width=0.24\linewidth]{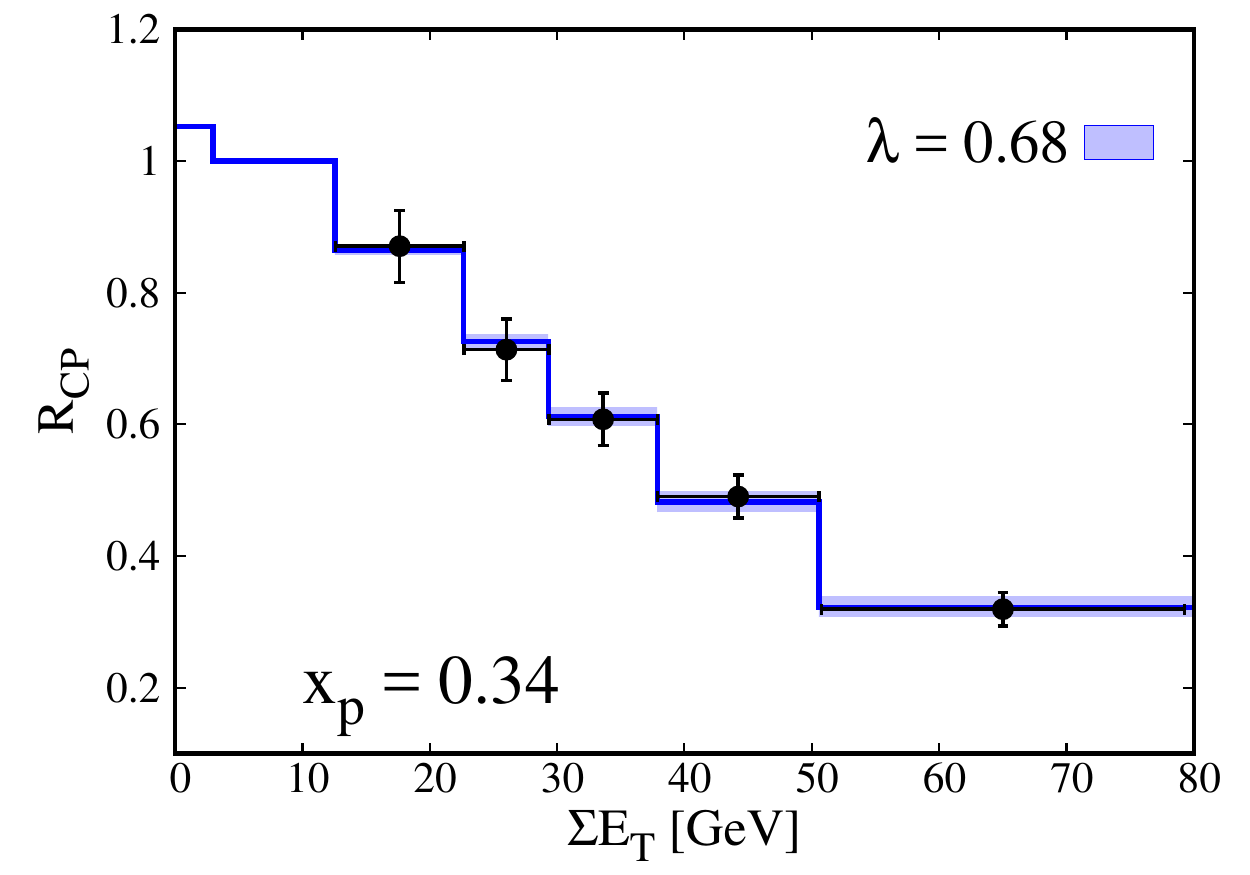}
\includegraphics[width=0.24\linewidth]{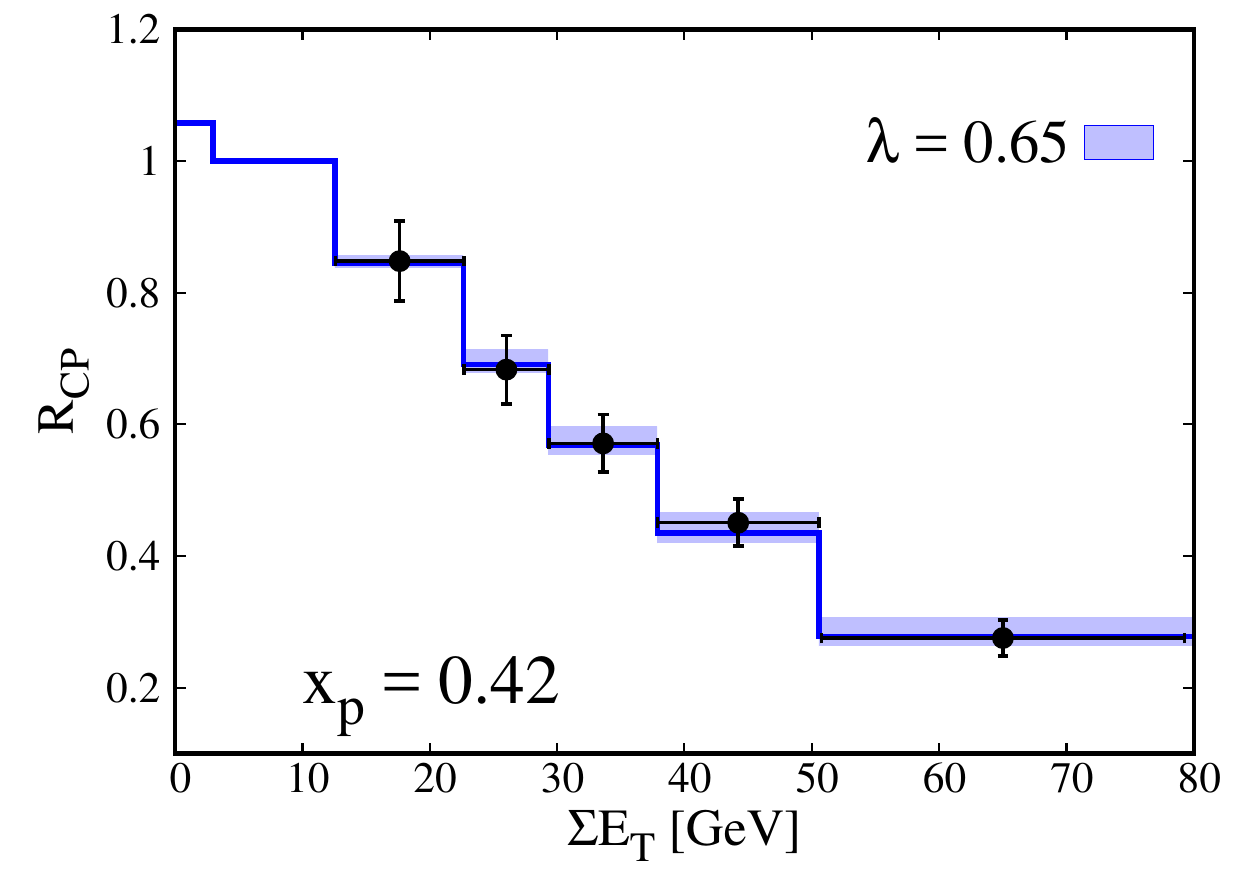}
\includegraphics[width=0.24\linewidth]{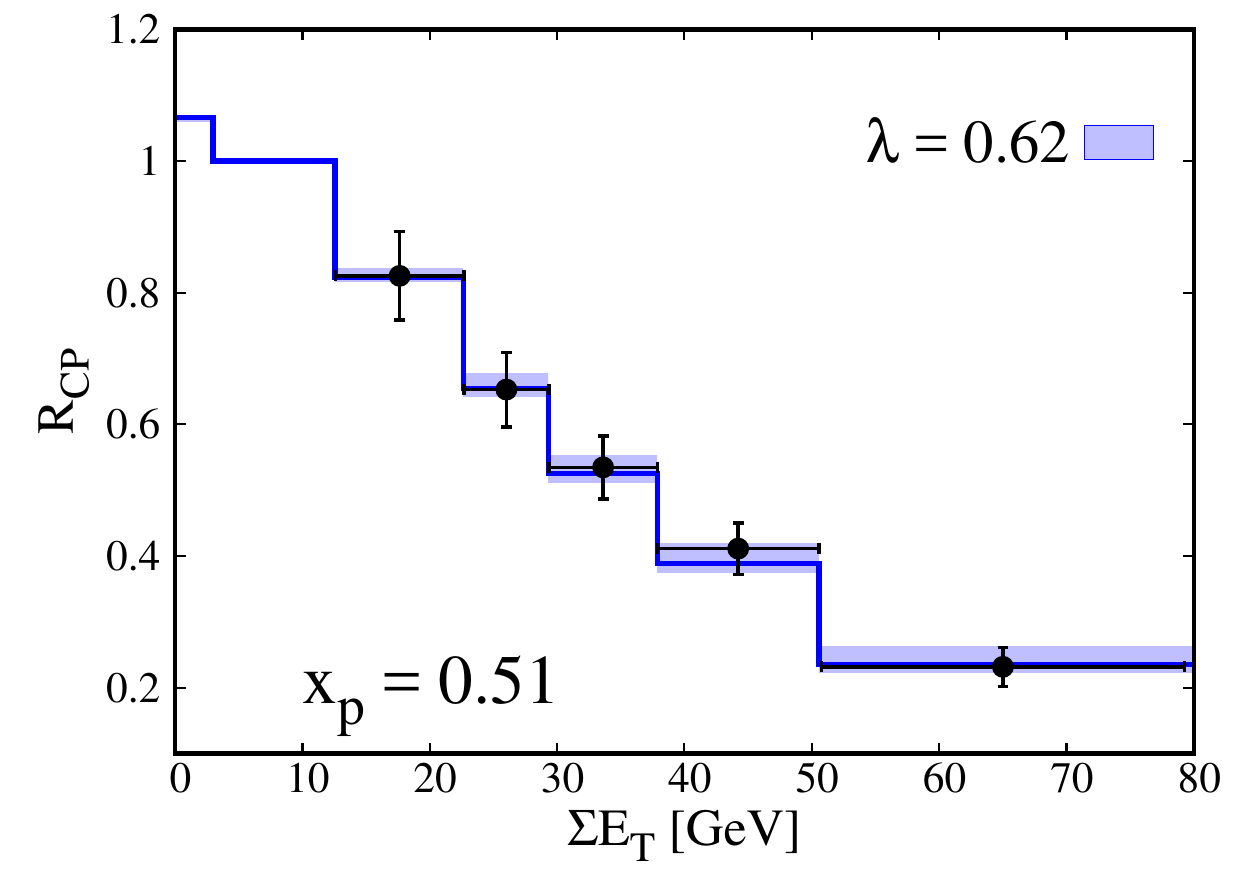}
\includegraphics[width=0.24\linewidth]{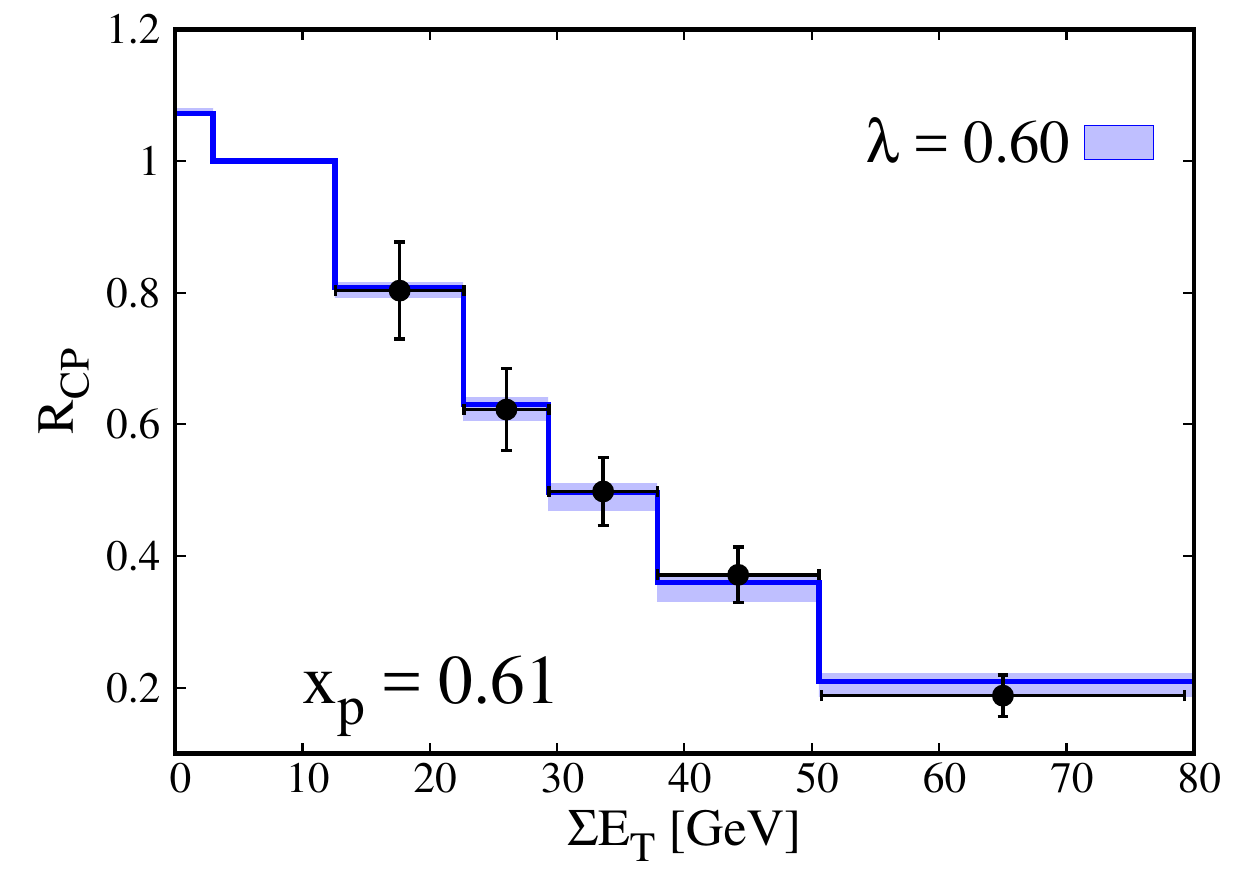}
\includegraphics[width=0.24\linewidth]{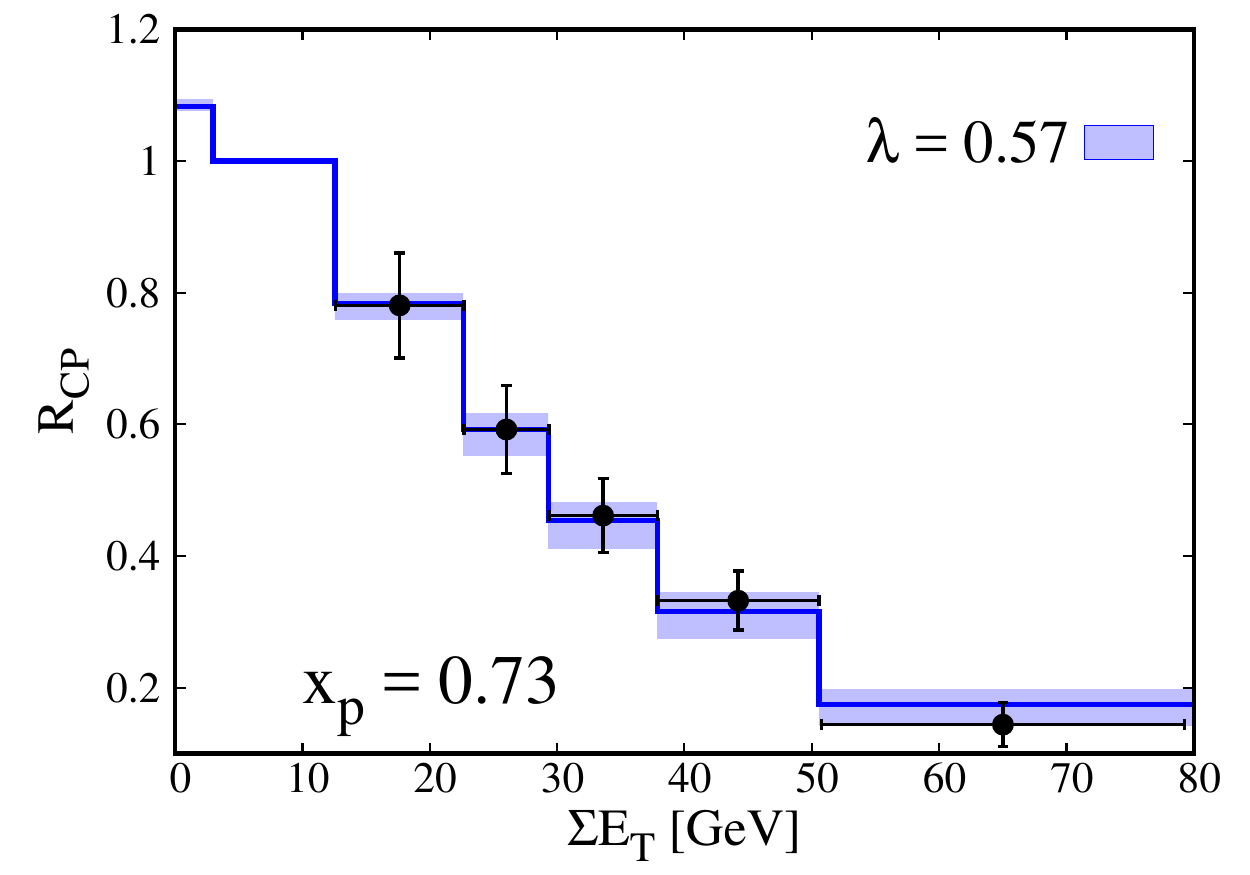}
\caption{\label{fig:fig3} Comparison of the LHC proton--lead nuclear
  modification factor data (black points) in different hadronic
  activity bins, to those in our model (shaded band). Each panel shows
  a different $x_p$ range.}
\end{figure*}

To explore how hard scattering rates are correlated with $\nu$, we
define the ratio

\begin{eqnarray}
R(\nu) = \left( \sigma^{hard}_\nu/\sigma^{MB}_\nu \right ) / \left(
\nu \cdot \sigma^{hard}_{NN}/ \sigma^{MB}_{NN} \right) \nonumber \\ = \left(
\sigma^{hard}_\nu/\sigma^{hard}_{NN} \right ) / \left( \nu \cdot
\sigma^{MB}_{\nu}/ \sigma^{MB}_{NN} \right),
\label{eq:hard}
\end{eqnarray}

\noindent where $\sigma^{hard}_\nu$ and $\sigma^{hard}_{NN}$ are the
hard process cross-section in $p$+A collisions with $\nu$ $NN$
interactions and just in one $NN$ collision, respectively, and
$\sigma^{MB}_{\nu}$ and $\sigma^{MB}_{NN}$ are the analogues of these
but for minimum bias (inelastic) collisions. $R(\nu)$ is the ratio of
the observed hard process rate to the rate expected given the number
of (soft) inelastic $NN$ interactions. Hence, the experiments observed
$R > 1$ for small $\nu$, $R < 1$ for large $\nu$, and $R=1$ for
$\nu$-integrated collisions.

We define the $x_p$-dependent shrinking of the average interaction
strength at a given collision energy $\sqrt{s}$ as

\begin{equation}
\lambda(x_p) = \left<\sigma^{MB}_{NN}(x_p)\right> / \sigma^{MB}_{NN}.
\end{equation}

The distribution over the number of collisions is mainly sensitive to
the value of $\lambda(x_p)$.  It has a small sensitivity to the size
of the fluctuations of $\sigma_{NN}(x_p)$. Hence, similar to what was
done in Ref.~\cite{Alvioli:2014eda}, we model fluctuations in the
strength of interaction at fixed $x_p$ by assuming that the dispersion
of $\sigma$ at fixed $x_p$ is similar to the average dispersion. As
$\lambda(x_p)$ decreases from unity, the deviations of $R(\nu)$ from
unity smoothly increase. For a given value of $\lambda(x_p)$, our
model provides $R(\nu)$ for each $\nu$.

The value of $R(\nu)$ is schematically identical to the experimentally
measured nuclear modification factors $R_\mathrm{pA}$
($R_{p\mathrm{Pb}}$ or $R_{d\mathrm{Au}}$), except that these are
reported for different {\em centrality} selections: sets of events
experimentally characterized by some range of hadronic activity at
large nuclear-going rapidity. In $p$+Pb collisions in
ATLAS~\cite{Aad:2015zza}, the hadronic activity is measured as the
transverse energy sum, $\Sigma{E}_\mathrm{T}$, in the hadronic
calorimeter situated at $-4.9 < \eta < -3.2$, and is taken to be
proportional to $\nu+1$ (the total number of participating
nucleons). In $d$+Au collisions in PHENIX~\cite{Adare:2015gla}, the
hadronic activity is defined as the total charge measured in the
beam--beam counter situated at $-3.9 < \eta < -3.1$, and is taken to
be proportional to $\nu$. In both cases, the selected hadronic event
activity (i.e. centrality) ranges result in sets of events with broad
but well-separated distributions of $\nu$. To compare our model with
the LHC and RHIC jet production data, we use the relationships between
$\nu$ and $\Sigma{E}_\mathrm{T}$ or charge established by the
experiments in Refs.~\cite{Adare:2013nff,Aad:2015zza} to determine the
distributions over $\nu$ for each centrality selection. Thus, for each
value of $\lambda(x_p)$, we calculate the nuclear modification
factors, $R_\mathrm{pA}$, weighted by the $\nu$ distribution in each
experimentally defined centrality selection.

Based on this model, we fit the ATLAS and PHENIX data in every bin of
$x_p \approx 2 p_t \cosh(y) / \sqrt{s}$ reported in the experiments to
find the best value of $\lambda(x_p)$ which describes $R_\mathrm{pA}$
in all reported centrality selections. In both datasets, we compare to
the so-called central-to-peripheral ratio, $R_{CP}$, which is the
ratio of $R_\mathrm{pA}$ in a given central event selection to that in
the most peripheral one. Since the centrality-averaged $R_\mathrm{pA}$
values are consistent with unity, the $R_{CP}$ values encode the same
information on the centrality dependence but with improved
experimental uncertainties for our fits.

We determine the best $\lambda(x_p)$ by minimizing the $\chi^2$ summed
over all centrality selections $i$, $\chi^2 = \sum_{i}
\left(R^{data}_i - R^{model}_{i}(\lambda) \right)^2 / \epsilon^2_{i}$
where $\epsilon^{2}$ is taken to be the quadrature sum of the
statistical and systematic uncertainties in the data. The RHIC and LHC
data provide three and five centralities for each value of $x_p$,
which are used to fit a single value of $\lambda(x_p)$, and they
provide data on eight and ten values of $x_p$ in total. In each $x_p$
range, we estimate the uncertainty on the extracted value of
$\lambda(x_p)$ as the range over which the $\chi^2$ increases by
one.

We note that there may be additional uncertainties in the modeling of
$P(\sigma, s)$, such as the variance of the distribution. These arise
from the lack of appropriate diffractive $pp$ data at RHIC and LHC
energies, and are thus difficult to quantify. However, the reasonable
agreement of the model with the data obtained at very different
energies and kinematic selections below suggests that the observables
considered here have only a moderate sensitivity to these details.

Figs.~\ref{fig:fig2} and~\ref{fig:fig3} show the full comparison of
the predictions of our model to RHIC and LHC data, respectively.

\begin{figure}[!t]
\includegraphics[width=1.0\linewidth]{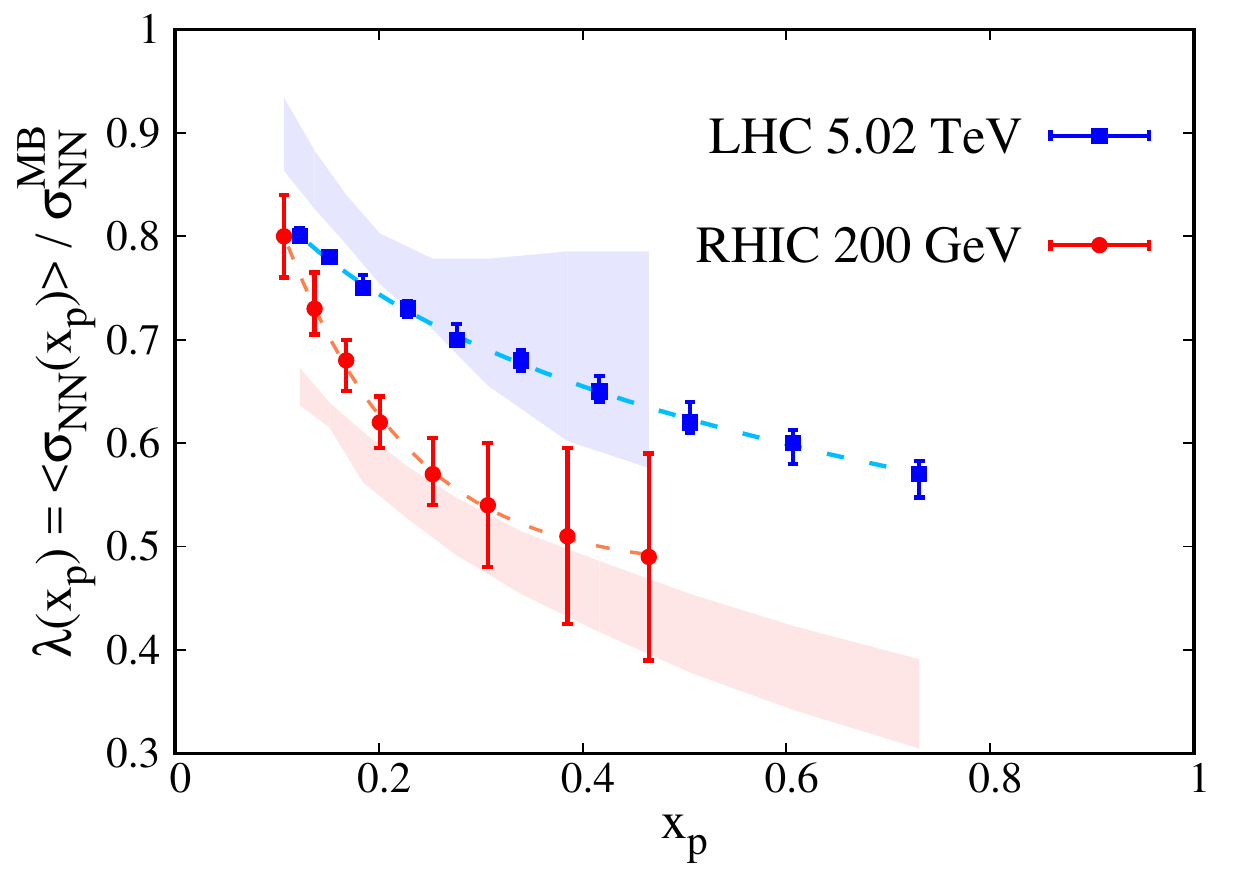}
\caption{\label{fig:fig4} Extracted values of $\lambda(x_p)$ as a
  function of $x_p$ at RHIC and LHC energies (solid points), with fits
  to an exponential function in $x_p$ shown as dashed lines to guide
  the eye. The shaded bands are a prediction for $\lambda(x_p)$ at
  each energy using the results at the other energy as input (see
  text).}
\end{figure}

Fig.~\ref{fig:fig4} summarizes the results of our global analysis of
$\lambda(x_p)$ as a function of $x_p$ and collision energy. In the
case of the RHIC data, our analysis yields slightly smaller values of
$\lambda(x_p)$ than those in Ref.~\cite{McGlinchey:2016ssj}, due to
differences in the treatment of the collision geometry. At low values
of $x_p \sim 0.1$, $\lambda(x_p)$ is similar at both RHIC and LHC
energies. At increasingly larger $x_p$, $\lambda(x_p)$ systematically
decreases but does so faster at RHIC energies.

These findings verify our previous expectations in
Ref.~\cite{Alvioli:2014eda} and have a natural explanation. In
perturbative QCD the total cross section for a bound state with a
small transverse size $\rho$ to interact with a nucleon is
proportional to the gluon density $g(Q^2, x_p)$ in the nucleon at
resolution scales $Q^2\propto 1/\rho$ and $x_p\sim Q^2/s$. At large
$Q^2$, $g$ grows quickly with decreasing $x_p$, resulting in an
increase of the cross-section (and of $\lambda(x_p)$ at fixed $x_p$)
for these small configurations with increasing collision
energy. However, this increase is slower than what is observed for
perturbative processes with vacuum exchange in t-channel, such such as
$J/\psi$ exclusive photoproduction~\cite{Abramowicz:1998ii}. Thus the
interaction at high energies may be thought of as lying between the
perturbative and non-perturbative domains, suggesting that chiral
symmetry is restored for the probed components of the light cone
proton wave function. Finally, the fast growth of the cross section
for small configurations is consistent with the expected narrowing of
the $P_N(\sigma)$ distribution at increasing collision
energies~\cite{Guzey:2005tk}.

A consistency check of our results can be performed under the
assumption that the probability to find a configuration with some
large $x_p$ is the same at two collision energies $\sqrt{s_1}$ and
$\sqrt{s_2}$. If the fluctuations in $\sigma(x_p)$ are small such
that, at fixed $x_p$, there is a one-to-one correspondence between
$\sigma(x_p)$ at two different energies, one may express this as the
probability to find a configuration with cross section smaller than
$\lambda(x_p)\sigma_{tot}$,

\begin{equation}
  \int_0^{\lambda(x_p;
    \sqrt{s_1})\sigma_{tot}(\sqrt{s_1})}\hspace{-1.5cm}d\sigma\,P_N(\sigma;
  \sqrt{s_1})\,=
  \,\int_0^{\lambda(x_p;\sqrt{s_2})\sigma_{tot}(\sqrt{s_2})}\hspace{-1.5cm}d\sigma\,P_N(\sigma;
  \sqrt{s_2}),
  \label{eq:energy}
\end{equation}

\noindent which along with Eq.~(\ref{psigma}) is an implicit equation
for the energy dependence of $\lambda(x_p)$ at fixed $x_p$.

Starting with the LHC results for $\lambda(x_p)$, we use
Eq.~\ref{eq:energy} to systematically predict $\lambda(x_p)$ at RHIC
energies at the same values of $x_p$, and vice
versa. Fig.~\ref{fig:fig4} shows the results of this check. For $x_p
\gtrsim 0.15$, the relationship between the extracted $\lambda(x_p)$
values at RHIC and LHC energies is consistent with that predicted by
Eq.~\ref{eq:energy}. At lower $x_p$, this method predicts a larger
difference in $\lambda(x_p)$ at the two energies than is extracted in
data, suggesting that our model does not provide a complete
description of color fluctuation phenomena in this $x_p$ range (for
example, since it ignores a possible parton flavor dependence). Using
the parameterization for $P_N(\sigma)$ at the lower, fixed--target
energies given in Ref.~\cite{Blaettel:1993rd}, one finds that
$\lambda(x_p \sim 0.5) \approx 0.38$ at $\sqrt{s} = 30$~GeV. At these
lower energies, the large-$x_p$ quarks are thus localized in an area
of transverse size $\sqrt{\lambda(x_p)} \approx 0.6$ smaller than that
in the average configuration, leading to them having a significantly
larger nonperturbative transverse momentum.


Recently, data on $200$~GeV proton--gold collisions were recorded at
RHIC, allowing for a further test of our model. Using the same
parameters which relate $\nu$ to the hadronic activity as in the
$d$+Au data, we calculate the distributions of $\nu$ in example
centrality bins and the $R_{CP}$ values for hard triggers with
different ranges of $x_p$. These predictions are summarized in
Fig.~\ref{fig:fig2}. As also argued in Ref.~\cite{McGlinchey:2016ssj},
the magnitude of the observable effect should be larger than in the
$d$+Au data, where it is expected to be washed out by the additional
projectile nucleon.

The global analysis presented in this paper quantitatively extends our
initial interpretation of the LHC data on forward jet production in
$p$+A collisions as arising from an $x_p$-dependent decrease in the
interaction strength of proton configurations~\cite{Alvioli:2014eda},
and demonstrates that the same picture successfully describes RHIC
data on large-$x_p$ jet production. Our analysis finds that the
suppression of the interaction strength is stronger at lower energies,
consistent with expectations from QCD that cross-sections for small
configurations grow faster with energy than do those for average
configurations.

Measurements of other processes arising from a different mixture of
large-$x_p$ quarks and gluons (e.g. Drell-Yan or electroweak
processes) would allow for a comparison of quark- vs. gluon-dominated
configurations. Analogous studies in ultraperipheral collision
data~\cite{Alvioli:2016gfo} may probe color fluctuations in the photon
wave function.

Our conclusions also have implications for understanding features in
the quark--gluon structure of nuclei such as the observed suppression
of the nuclear structure function at large-$x$, commonly known as the
EMC effect~\cite{Hen:2016kwk}. Since nucleons in a configuration with
a large-$x$ parton are weakly interacting and the strength of the
interaction at fixed $x$ falls at lower energies, it is natural to
expect that such configurations interact very weakly with other
nucleons at the energy ranges relevant for nuclei. In the bound
nucleon wavefunction, such weakly interacting nucleon configurations
are strongly suppressed~\cite{Frankfurt:1985cv}. Thus, this picture
suggests a natural explanation for the observed suppression of partons
in the EMC effect region. This phenomenon may furthermore provide
information on how the properties of nucleons experiencing large
pressures may change, leading to, for example, the restoration of
chiral symmetry within the core of neutron stars.


\begin{acknowledgments}

We thank B. Muller for the suggestion to present predictions for $p$+A
running at RHIC within our framework, A. Mueller for discussion of
proton squeezing at large $x_p$, and J. Nagle for suggestions on the
manuscript. L.F.'s and M.S.'s research was supported by the US
Department of Energy Office of Science, Office of Nuclear Physics
under Award No. DE-FG02-93ER40771. D.V.P.'s research was supposed by
the US Department of Energy Office of Science, Office of Nuclear
Physics under Award No. DE-SC0018117.

\end{acknowledgments}

\bibliography{ColorFluctuationPaper}   

\end{document}